\documentclass{article}

\usepackage{PRIMEarxiv}

\usepackage{amsmath}
\usepackage[utf8]{inputenc} 
\usepackage[T1]{fontenc}    
\usepackage{hyperref}       
\usepackage{url}            
\usepackage{booktabs}       
\usepackage{amsfonts}       
\usepackage{nicefrac}       
\usepackage{microtype}      
\usepackage{lipsum}
\usepackage{fancyhdr}       
\usepackage{graphicx}       
\graphicspath{{media/}}     

\title{The Collective Burst Mechanism of Angular Jumps in Liquid Water
\thanks{\textit{\underline{Citation}}: 
\textbf{Authors. Title. Pages.... DOI:000000/11111.}} 
}

\author{
Adu Offei-Danso$^{a,b}$, Uriel N. Morzan$^{a}$, Alex Rodriguez$^{a}$, Ali Hassanali$^{a}$, Asja Jelic$^{a}$\\
$^{a}$The Abdus Salam International Centre for Theoretical Physics, 34151 Trieste, Italy\\
$^{b}$International School for Advanced Studies (SISSA), 34136 Trieste, Italy\\
}

\begin{document}
\maketitle

\begin{abstract}

Understanding the microscopic origins of collective reorientational motions in aqueous systems requires techniques that allow us to reach beyond our chemical imagination. Herein, we elucidate a mechanism using unsupervised learning, showing that large angular jumps in liquid water involve highly cooperative orchestrated motions. Our automatized detection of angular fluctuations, unravels a  heterogeneity in the type of angular jumps occurring concertedly in the system. We show that large orientational motions require a highly collective dynamic process involving correlated motion of up to 10\% of water molecules in the hydrogen-bond network that form spatially connected clusters.
This phenomenon is rooted in the collective fluctuations of the network topology which results in the creation of defects in waves on the ThZ timescale.
The mechanism we propose involves a cascade of hydrogen-bond fluctuations underlying angular jumps and provides new insights into the current localized picture of angular jumps, and in its wide use in the interpretations of numerous spectroscopies as well in reorientational dynamics of water near biological and inorganic systems.
\end{abstract}


\section{Introduction}
Hydrogen-bond (H-bond) network fluctuations in water are at the heart of a wide range of physical, chemical, and biological processes, ranging from proton transfer in the ionization of water \cite{ando1995hf,hassanali2013proton} to the folding of proteins and aggregation of molecules in solution \cite{jong2018data,laage2009water}. 
Since water molecules are characterized by a rather large dipole moment, which in turn leads to directed interactions between them, the molecular reorientation mechanism has attracted significant interest \cite{stirnemann2012communication,biswas2017ir,ramasesha2011ultrafast,nicodemus2011collective}. 

Over a decade ago, Laage and Hynes demonstrated that water rotations do not occur solely via small diffusive steps, but instead typically involve sudden large-amplitude angular jumps \cite{laage2006molecular,laage&hynes,laage2009reinterpretation,laage2012}. The mechanism by which large and quick rotations of a water molecule happen is due to fluctuations in the local coordination patterns of neighbouring waters suggesting a process that involves at least three water molecules which alters the local H-bond network topology. When these jumps occur, do other water molecules in this network remain as spectators or are they active participants in a more collective process?

One of the enormous hurdles in determining a clear answer to this question originates from the difficulty in disentangling fluctuations occurring over a wide spread of both length and timescales, that create and form labile hydrogen bonds with a plethora of hydrogen-bond network patterns. On the other hand, numerous experiments ranging from dielectric spectroscopy\cite{popov2016mechanism} to time-dependent vibrational spectroscopy\cite{fecko2003ultrafast,fecko2005local,nicodemus2011collective,biswas2017ir,ramasesha2011ultrafast}. indicate the presence of collective and concerted processes\cite{heyden2010dissecting,ramasesha2011ultrafast,moilanen2009ion,nicodemus2011collective,stenger2002photon} underlying reorientational dynamics in water.

Here, using an automatized detection of angular motions determined from molecular dynamics simulations of water, we illustrate a mechanism that unambiguously reveals the collective and correlated nature of water reorientation dynamics. By using an unsupervised protocol that identifies all abrupt angular motions, with no \emph{a priori} assumptions on hydrogen-bonding, we demonstrate that there is a heterogeneity in the types of angular jumps that occur and that large reorientations are facilitated by a highly orchestrated motion of dozens of water molecules. We assert that these features are a generic property of the fluctuations in the topology of the water H-bond network on the THz timescale. We also show that regions with lower local density serve as hot-spot sites in the network where large reorientations can occur simultaneously.

\section{Materials and Methods}
\label{sec:headings}

\subsection*{Molecular Dynamics} 

We performed a molecular dynamics simulation of 1019 water molecules using the GROMACS 5.0 package \cite{bekker1993gromacs} with the SPC-E rigid water model \cite{berendsen1987missing}. This water model was used previously by Laage and Hynes\cite{laage&hynes}. 
Energy minimization was first carried out to relax the system, followed by an NPT and subsequently NVT equilibration at 300K and 1 atmosphere for 10ns each. A timestep of 1fs is used for all the simulations. The NVT simulations were performed using the canonical velocity-rescaling thermostat \cite{bussi2007canonical} with a time constant of 2ps. The choice of the thermostat also ensures that dynamical properties of water are not disturbed. The NPT runs were conducted using the Parrinello-Rahman \cite{parrinello1980crystal} barostat with a pressure coupling time constant of 2ps. The production run at 300K was carried out for 2ns in the NVT ensemble where the trajectory was outputted every 4fs in order to resolve the dynamics associated with the fast reorientations. The dimensions of the cubic box used for the NVT simulations was 32\AA{}. In the Supplementary Information (SI), we also compare some of our results with the MB-pol potential which is among the most accurate in-silico potentials reproducing many of the properties of water across the phase diagram \cite{babin2013development,reddy2016accuracy}.

\subsection*{Unsupervised detection of angular swings}

Water reorientation dynamics includes various processes happening at different time scales, from very fast vibrational motions causing limited reorientation, to slower reorientation through sudden large-amplitude angular jumps. 
It remains still an open question how and to which extent each of these processes is involved in the underlying collective hydrogen bond rearrangements.
To elucidate the mechanism behind the collective reorganization of water observed in Fig.\ref{fig1}, we developed an automatized protocol for detecting all the various angular changes in water reorientation, which we term angular swings. Our protocol identifies these angular swings without any prior knowledge nor by using geometric or energetic criteria for the hydrogen bonding interactions.
Here, we describe all the steps of the protocol depicted in Fig.\ref{protocol2}.

In order to track down angular changes in water orientation, we rely on two body-fixed vectors, the dipole moment (DP) and the HH vector (see Fig.\ref{protocol2}A). 
From the MD simulation, we first extracted the DP and HH vector time series for each water molecule, herein referred to as $\vec{v}(t)$, with $t$ being time. A new times series, $\vec{v}_{\text F}(t)$, was  constructed by filtering $\vec{v}(t)$ using a second-order low-pass digital butterworth filter \cite{butterworth1930theory} implemented in MATLAB \cite{MATLAB2020a} with the cutoff frequency of 25 THz. To smoothen the time series, a mean filter of 
10THz 
was then subsequently applied. An example of the original unfiltered and the filtered time series of one of the DP vector components are shown in Fig.\ref{protocol2}A, where we can observe very clear sudden changes, such is the one at around time 1500 fs, that should be detected by the automatized protocol.

Using the filtered time series we subsequently compute the derivative of $\vec{v}_{\text F}(t)$ using a finite difference method, and calculate the cross product of this vector with 
$\vec{v}_{\text F}(t)$ in order to obtain a new vector
\begin{equation}
    \vec{n}(t)= \vec{v}_{\text F}(t) \times \frac{d \vec{v}_{\text F}(t)}{d t},
\end{equation}
which corresponds to the vector perpendicular to the plane of rotation of the body-fixed vector $\vec{v}_{\text F}(t)$. 
Our protocol defines an angular swing to be a process that does not change the plane of rotation of the body-fixed vector $\vec{v}_{\text F}(t)$. 
This implies that, over the time of one angular swing, the direction of $\vec{n}(t)$ does not change, as shown in Fig.\ref{protocol2}B. The start and the end points of swing events are then identified as large instantaneous changes in the direction of $\vec{n}(t)$.  
More precisely, we look at the following quantity
\begin{equation}
 q(t) =1 - \frac{\vec{n}(t) \cdot \vec{n}(t+dt )}{|\vec{n}(t) || \vec{n}(t+dt)|},
\end{equation}
which is equal to $0$ during the swing. At the start and at the end of the swing, this quantity is found to be non-zero, indicating the change in the plane of rotation of the body-fixed vector $\vec{v}_{\text F}(t)$. Consequently, the start and end times of angular swings can be identified as maxima in $q(t)$,
which are fully consistent with extrema in the filtered time series, $\vec{v}_{\text F}(t)$, correctly recognizing sudden angular changes as shown in Fig.\ref{protocol2}A.  

Finally, having identified the start and end points of the swings, the duration of the swings are taken to be the times between two peaks of $q(t)$, and the magnitude is found by computing the angle between the unfiltered DP or HH vector at the start and at the end point of the swing.
The final output of the protocol is the start time (t), duration ($\Delta t$), and magnitude $(\Delta \Theta)$, for each angular swing detected. We perform the procedure both for DP and HH vectors, as some angular fluctuations of water molecules can be better captured through one or the other vector.

\section*{Results and Discussion}

\begin{figure*}[!htb]
\centering
\includegraphics[width=\textwidth]{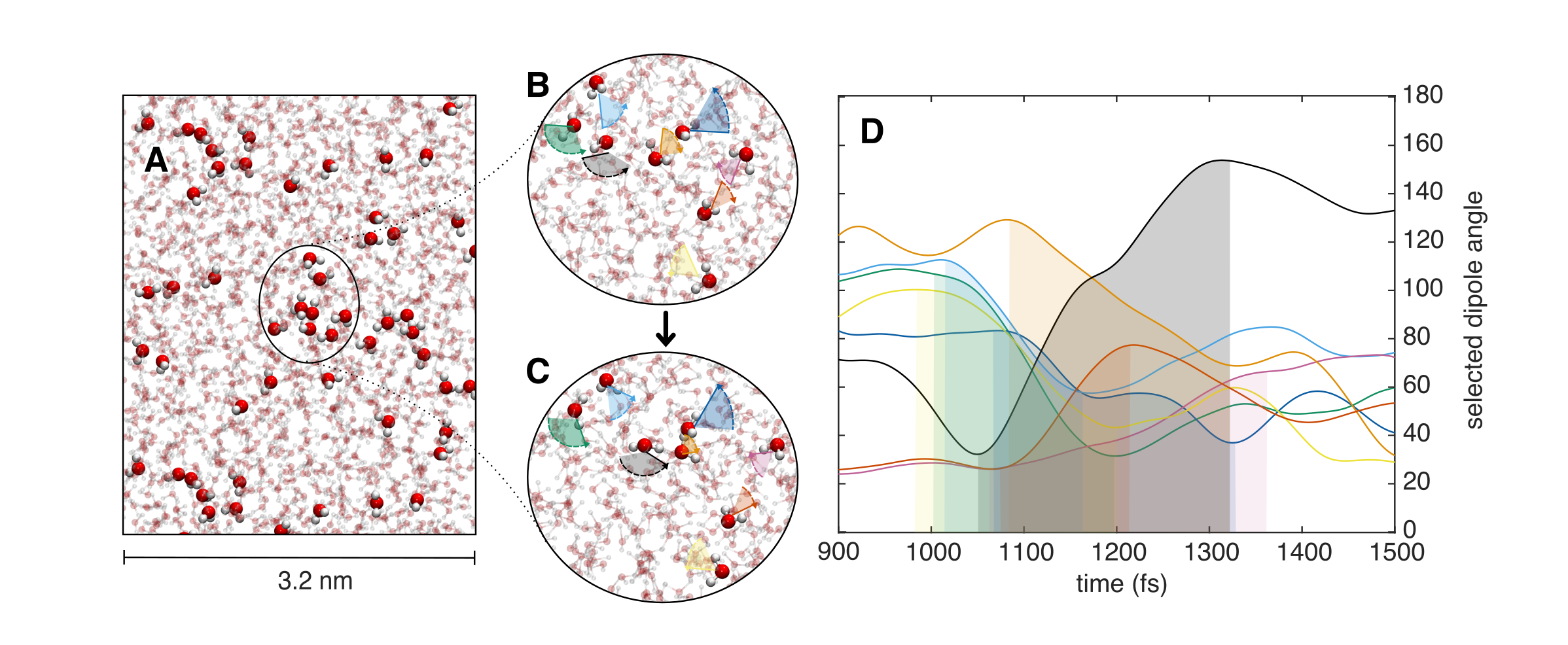}
 \caption{Collective nature of angular jumps. 
 (A) Highlighted are all molecules undergoing angular reorientation of magnitude greater than 60 degrees in a box of 3.2nm within the time interval of 350fs (which spans between time steps 1000fs and 1350fs in the MD simulation). They amount to around 5\% of the total number of 1019 simulated molecules.
 (B,C) Close-ups of 8 of these molecules in a small region of the box at the start (panel B) and at the end (panel C) of a large angular jump as observed from the changes in their dipole vectors. The colored arcs outline the angular motion carried by the dipole vectors in the direction of the dashed arrow. Positions of the molecules in panels B and C are slightly different due to translational motion during the observed time interval.
(D) 
Change of the dipole vector in time for each of the selected molecules: We plot 
the time evolution of the angle it forms with respect to one of the axes of the laboratory coordinate system (for each molecule we show the component which changes most in this time interval). 
The regions between the start and the end of the angular jump are shaded by the colours of the corresponding molecules in panels B and C.
}
\label{fig1}
\end{figure*} 

In order to build our intuition on the collective nature of angular jumps, in Fig.\ref{fig1}A we highlight a specific event captured by our protocol illustrating all water molecules in the system that perform a large-amplitude angular reorientation within a selected time interval of approximately 350 fs in the course of the molecular dynamics (MD) simulation. From the total of 1019 water molecules, we find that around 5\% of them undergo a major change in the direction of their dipole moment vector during this short time interval, suggesting that these are concurrent events. Many of these molecules are also located close by in space, forming apparent clusters of large jumping molecules. 
The round panels in the middle (Fig.\ref{fig1}B,C), show a close-up of several of these molecules before and after the angular jump, which is made evident by their initial and final dipole vector orientations. 
Shown in the background are all the other water molecules in close vicinity to this event. For each reorienting molecule in the selected group, we quantified its angular fluctuation in Fig.\ref{fig1}D by following how its dipole vector changes in time through the time evolution of the angle it forms with respect to one of the axes of the laboratory coordinate system. 
While this only serves as a proxy for the angular fluctuations, it gives us a semi-quantitative measure of the size of the angular reorientation.

We see that the angular change of the dipole vectors of the eight molecules involved in this event, ranges between $60$--$120$ degrees within the selected time interval. This type of angular change clearly modifies the direction in which the dipoles of these eight water molecules point to. Moreover, eye-balling the time series shown, indicates that these reorientation events are simultaneous and possibly correlated in space, which, as we will see shortly, requires a collective reorganization of the topology of the hydrogen-bond network. All this suggests that there is a highly coordinated dynamics underlying angular jumps, involving large number of molecules and diverse motions, which will be elucidated in the ensuing analysis.

\begin{figure*}[!htb]
\centering
\includegraphics[width=\textwidth]{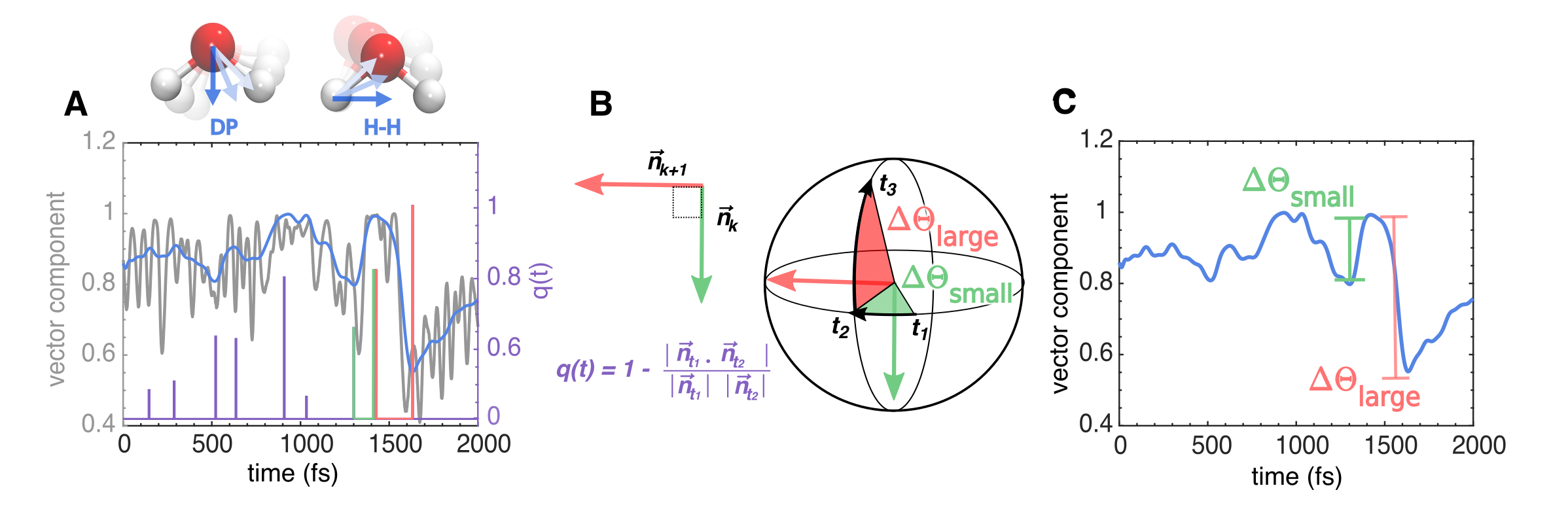}
\caption{Angular swing detection protocol. 
(A, upper) Definition of the dipole (DP) and HH vectors, obtained for every molecule in the system. (A, lower) The plot shows the time series of the DP vector's x-component (DP$_{\text{x}}$) for a selected water molecule in a 2000 fs interval (gray line). The blue curve corresponds to the filtered time series eliminating the high frequency oscillations in the DP$_{\text{x}}$ vector. The purple line correspond to the function $q(t)$ defined in the Methods section that presents a spike at the beginning and the end of every angular swing. 
(B) Two successive swing events are detected by identifying the start and the end points of each angular swing. By employing function $q(t)$, we detect the instantaneous change in the direction of vector $\vec{n}(t)$
perpendicular to the plane of rotation of the DP (or HH) vector. This example shows our protocol for two successive swing events k-th and k+1-th in the time intervals $(t_{1},t_{2})$ and $(t_{2},t_{3})$, colored in green and red, respectively, detected from the filtered time series of the DP vector of one molecule. The direction of the vector normal to the plane of rotation is found to change only when transitioning from  one interval to the other, i.e.\ from $\vec{n}_{k}$ (green arrow) during $(t_{1},t_{2})$ time interval, to $\vec{n}_{k+1}$ (red arrow) during time interval $(t_{2},t_{3})$.  
(C) Angular swings $\Delta\theta_{\text{small}}$ (small green swing) followed by $\Delta\theta_{\text{large}}$ (large red swing) are indicated on the filtered DP vector component time series. The start and end points correspond to extrema in the time series. 
}
    \label{protocol2}
\end{figure*}

Quantifying the angular jumps illustrated in Fig.\ref{fig1}D requires the identification of reaction coordinates that are highly non-local, involving several degrees of freedom that are challenging to identify by eye. With the aim to investigate the mechanisms behind the putative collective reorganization of water from the MD simulations, we developed an automatized protocol for identifying all the various angular fluctuations in water reorientation. Illustration of the main steps of the procedure are shown in Fig.\ref{protocol2}, while more details are given in Materials and Methods. For each water molecule in the system, the procedure works by spotting sudden changes along the trajectories of the vectors defined within the molecular frame of a water molecule, namely the dipole moment (DP) vector and the HH vector. As an outcome of the automatized protocol, for each detected angular change we obtain its starting time, duration $(\Delta t)$, and angular magnitude $(\Delta\Theta)$. As we will see later, the angular changes we identified, have a broad range of angular amplitudes and duration consistent with previous observations of the local angular jump picture.
Therefore, we dubbbed all detected molecular reorientations as angular swings, rather than angular jumps. The latter is the predominant term in the literature, but it only refers to large-amplitude angular swings accompanied by H-bond breaking \cite{laage2006molecular,laage&hynes}. Here, we also analyze small angular fluctuations which would essentially correspond to small angular diffusive steps, that don't necessarily involve H-bond breaking. 

Underpinning the large angular jumps in the H-bond network, are fluctuations in the topology of water molecules. Large angular motions usually create coordination defects which affect the hydrogen bonding patterns \cite{laage&hynes}. This, in turn, affects local topology of the H-bond network, leading to rearrangements of nearby water molecules and possibly further large reorientation events. Here we examine the connection between changes in local topology and angular swings in more detail by quantifying the occurrence of these events in time for the entire ensemble of water molecules.

\begin{figure*}[!htb]
\centering
\includegraphics[width=\textwidth]{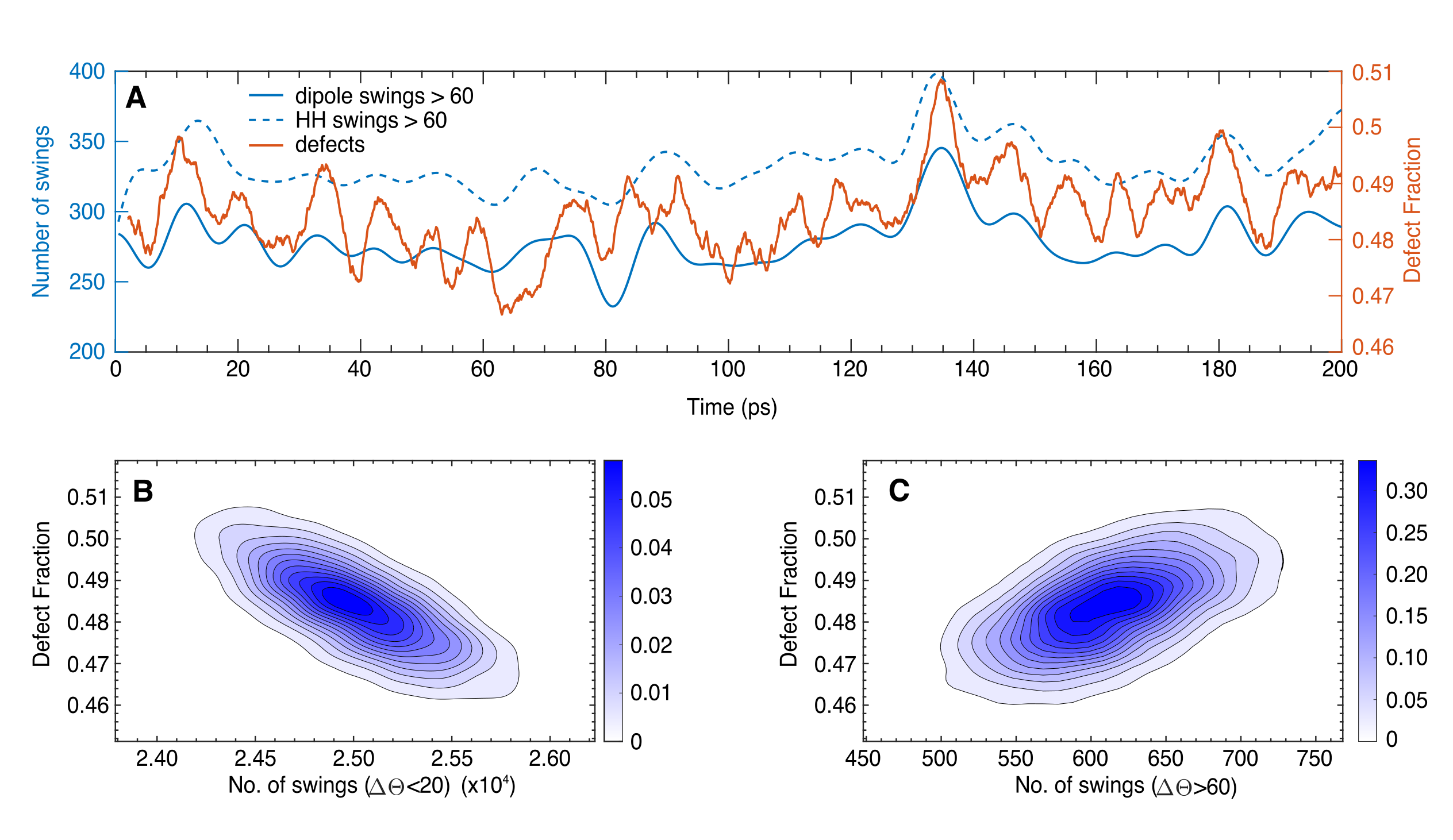}
\caption{Correlation between the number of simultaneous large angular swings and the fluctuations in the local topology of the water H-bond network.
(A) Time series of the number of molecules in the H-bond network performing large angular swings (amplitude larger than 60 degrees) at each moment of time as detected from the observation of the dipole vector (blue full line) and HH vector (blue dashed line). 
At each moment of time, we count the number of swings happening in the system within a time window of 1ps around it.
We superimpose these time series with the time series of the fraction of molecules in the H-bond network that are defective, i.e.\ with non-tetrahedral local topology (red). 
We observe fluctuations of the order of tens of picoseconds in all three curves that often appear to be correlated in time.
(B) Density plot of the fraction of defects in the H-bond network with respect to the number of molecules in the network performing small angular swings ($\Delta \Theta<20$) within 1ps. Anti-correlation between these two quantities means that when there are more molecules with defective local topology, the less small-amplitude angular swings occur in the H-bond network.
We find the correlation coefficient to be $-0.7390\pm0.0089$, with $p<0.01$. 
(C) Density plot of the fraction of defects in the H-bond network with respect to the number of large-amplitude angular swings ($\Delta\Theta>60$) within 1ps. 
Correlation between these two quantities indicates that the more the local topology in the H-bond network is defective, the larger is the number of molecules that perform large-amplitude angular swings.
The correlation coefficient found is $0.5604\pm0.0151$, with $p<0.01$.
}
\label{jumps_versus_defects}
\end{figure*}

First, at every time step of the simulation, we calculate the number of waters in the system that are non-defective, i.e.\ those that accept two and donate two hydrogen bonds, and the number of all the other water molecules, which we refer to as defects. 
The time series of the fraction of defects with respect to the total number of water molecules in the system is shown in Fig.\ref{jumps_versus_defects}A over a time interval of 200 ps  (see Fig.S1 for the time evolution of the fraction of defects over a longer time scale).
Interestingly, we observe that the fluctuations in the number of defective water molecules occur in waves. These oscillations reflect processes in the network which, on a picosecond timescale, for example, lead to the creation or annihilation of up to 10-20 defective water molecules in the network (large defect oscillations over short time scale are shown in Fig.S1). Many of these bursts then appear to accumulate over a longer timescale leading to a slower process occurring on 10s of picoseconds that is evident in Fig.\ref{jumps_versus_defects}A.

Large changes in the number of defects in the network suggests that there is a collective process underlying the water reorientations in the H-bond network. This phenomenon is akin to similar effects proposing the role of defects in the mobility of water under supercooling \cite{sciortino1991effect}.
To quantify better the collective nature of angular swings in the water network, we now look at how many angular swings occur simultaneously in the system.  In similar spirit to the idea of propagation of defects \cite{popov2016mechanism}, we assume that the rearrangements of the local H-bond network due to a molecular reorientation can trigger further swings close by in time.  Therefore, at every time step, we calculate how many molecules perform angular swings within a specific time interval around it.

Previous studies report that water reorientation occurs on a time scale of about 1ps that includes not only the angular jump itself, but also the breakage and forming of the H-bond before and after the jump \cite{laage2006molecular,laage&hynes}. 
Thus, we look at a time window of 1ps, within which we calculate all the angular swings happening in the system. 
By automatizing the detection of angular motions, we are able to identify all possible sudden angular changes in the waters' motion, spanning a wide range of duration and amplitude. 
However, large-amplitude swings are most likely the ones leading to water reorientation and change in local topology, as considered in the literature \cite{laage2006molecular,laage&hynes}. Moreover, the small-amplitude swings do not change much the orientation of the water molecule, and are less likely to affect local topology of the H-bond network. 

In Fig.\ref{jumps_versus_defects}A, we plot the number of swings with the angular amplitude larger than 60 degrees, detected from the angular motion of both DP and HH vectors. We find that the number of concurrent large swings in the system fluctuates on the timescale of 10s of picoseconds with the same frequency as the number of defective waters in the system.  Moreover, the oscillations seem to be correlated in time: the larger the number of the defected water molecules, the more large-amplitude swings simultaneously happening in the system. Although it has been well appreciated that a sizable fraction of the hydrogen-bond network involve the presence of coordination defects\cite{gasparotto2016}, their role in reorientational dynamics has remained elusive. In fact, in panels B and C of Fig.\ref{jumps_versus_defects}, we show that when the local topology in the H-bond network is more defective, there are less small-amplitude angular changes in the system and more of the large-amplitude ones. Fig. S2 in the Supporting Information illustrates similar analysis for swings with $\Delta\theta>40^\circ$.

Our findings suggest a deep connection between the fluctuations in the underlying topology of the H-bond network and the nature of angular swings occurring simultaneously in the system. These results reinforce a picture of water reorientations being an outcome of highly coordinated dynamics of water molecules, rooted in the collective fluctuations of the network's topology. Our findings bare some similarities with recent work by Liu and co-workers \cite{liu2013correlation,liu2018interplay} where they show that an angular jump of a given water molecule could enhance the subsequent jump motions of the same water molecule and surrounding water molecules up to the 2nd coordination shell.



Perhaps not surprisingly, when looking at how correlated in time the angular trajectories of two molecules performing large jumps \textit{concurrently} are, we find high correlation on the timescale of 10s of femtoseconds encompassing these angular jumps. However, as we look at the correlations between trajectories of the same molecules over longer times, they become less and less correlated akin to what is expected for any two random molecules in the system (see Fig.S3). While this confirms that there is indeed a high number of large jumps performed simultaneously by molecules, it still does not explain the collective nature and the mechanism underlying these events.

\begin{figure*}[!hbt]
\centering
\includegraphics[width=\textwidth]{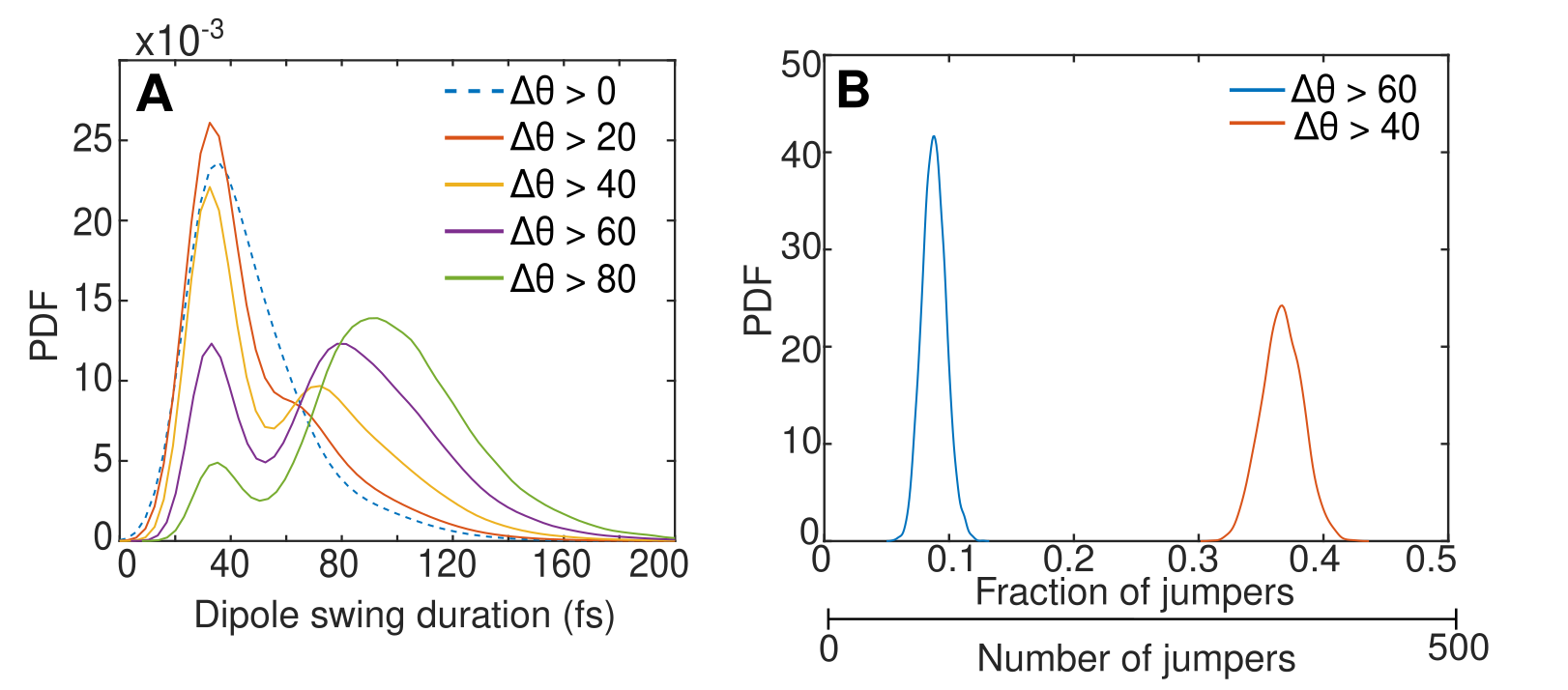}
\caption{
Duration and collectiveness of angular swings.
(A) Probability distributions of duration $\Delta t$ of swings with the angular magnitude $\Delta\Theta$ greater than a certain threshold, detected from the DP vector time series. 
(B) Probability distribution function of the number of molecules in the system that perform large angular swings within a time window of 200 fs, which are considered to be occurring concurrently. Average number of large swings with the magnitude $\Delta\Theta>60^{\circ}$ within the selected time window is around 100, which is around 10\% of the total number of molecules. Instead, the number of large reorientations with $\Delta\Theta>40^{\circ}$ is around 350, around 35\% of the total number of water molecules.
}
\label{fig:duration}
\end{figure*}

Clues into the origins of the collective reorientational dynamics can be seen in Fig.\ref{fig1}A which suggests that many of the highlighted molecules performing large jumps close by in time, also seem to be close by in space, as if forming clusters of concurrently jumping molecules. This suggests that underlying these neighboring large jumps there is a coordinated reorientational dynamics facilitated by the reorganization of the local H-bond network.  The specific details of the spatial distribution of the water molecules that simultaneously perform large angular jumps in the network will be tuned by both the timescales and magnitude of the reorientational dynamics. An important outcome of our automatized protocol illustrated earlier, is that it gives direct insights into this information through the statistics of $\Delta t$ as a function of $\Delta \theta$.

Fig.\ref{fig:duration}A shows the probability distributions for the duration of swings with the amplitude $\Delta\Theta$ larger than a certain angular threshold for the DP vector of the water molecules. For small angular thresholds, when the selected swings are predominantly of small magnitude, the swing duration peaks at roughly 30 fs.  This corresponds to a fast hindered rotational mode. As we increase the angular magnitude threshold, we find that the probability distributions change, both for the DP and the HH vectors (see Fig.S5A). For intermediate values of the angular amplitude, a second peak in the probability distribution appears and the distribution becomes bimodal. A new characteristic time of around 100 fs emerges when looking only at angular fluctuations with the magnitude larger than $60^{\circ}$, where the second peak in the bimodal distribution becomes prominent. This in turn corresponds to the characteristic time of large angular swings. Note that for the duration of the jump we consider just the angular reorientation of the water molecule, and not the time needed for the H-bond breakage and formation, as often considered in the literature \cite{laage2006molecular,laage&hynes}. The characteristic time of 100fs that emerges from our unsupervised protocol instead, corresponds more closely to the time it takes to transition from one stable hydrogen-bonded state to another. 


Armored with this information on the characteristic time, we first determined the total number of jumps that occur within a time interval of 200fs shifting the time window along the entire trajectory. Figure \ref{fig:duration}B shows the fraction of jumping water molecules in the ensemble obtained for two different $\Delta \theta$ thresholds. Interestingly, out of $\sim$1000 waters in our simulation box, our detection protocol shows that approximately 100 water molecules undergo jumps with $\Delta \theta>60^\circ$. As expected, relaxing the threshold of the magnitude of the jump to $40^\circ$ enhances the population undergoing angular transitions to $\sim$350 water molecules. If such a large number of water molecules undergo angular fluctuations, do they occur independently from each other or are there some underlying connections in the hydrogen bond network involved in these reorientational motions?

\begin{figure*}[!htb]
\centering
\includegraphics[width=\textwidth]{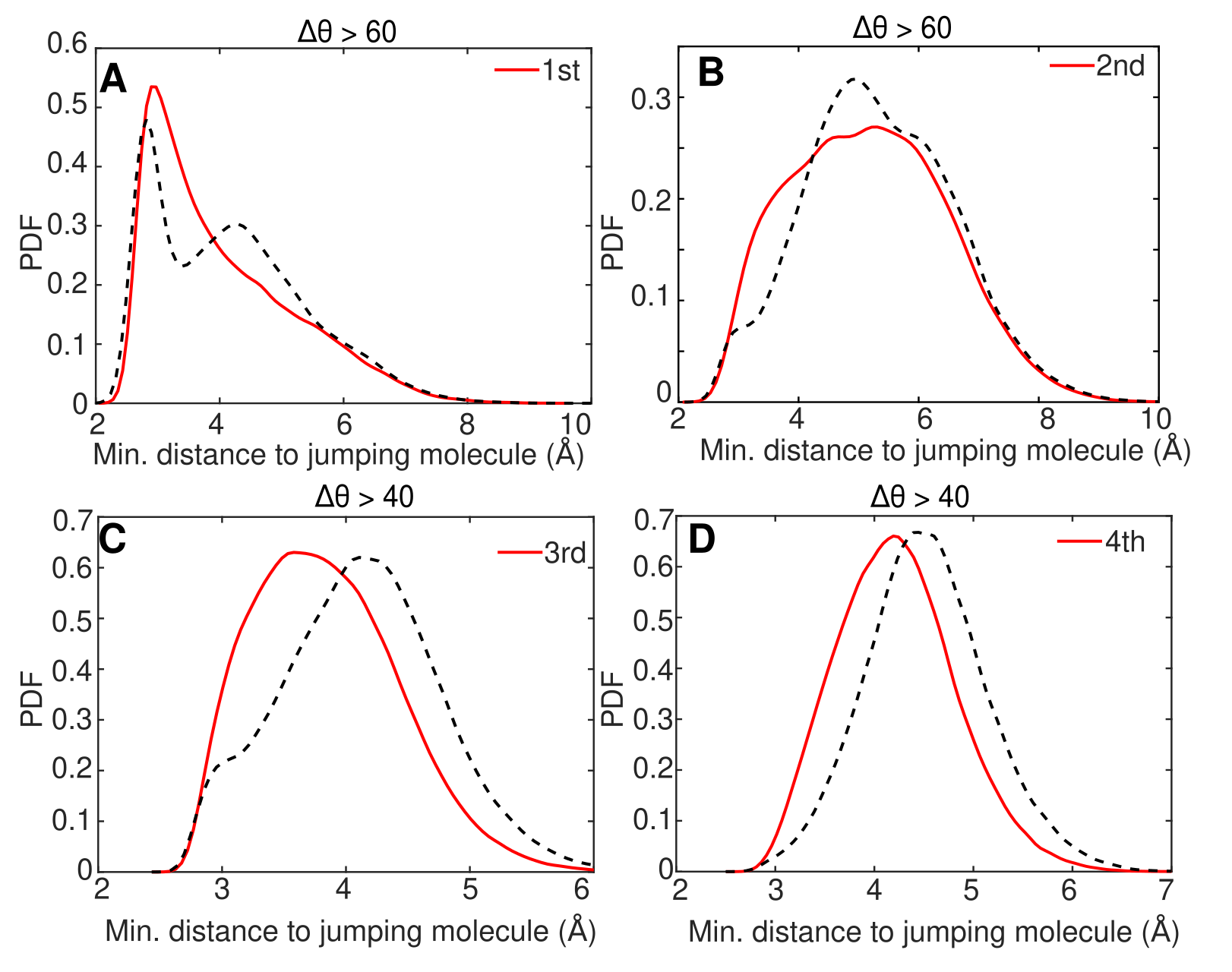}
\caption{
Large angular swings occur close by in space. 
(A-D) For the water molecules performing large swings within the selected time interval of 200 fs, full red lines show the probability distribution function of the distances to the first 4 nearest jumping molecules with respect to another jumping molecule. Large swings with amplitudes $\Delta\Theta>60^{\circ}$ and $\Delta\Theta>40^{\circ}$, are considered in panels (A,B) and (C,D), respectively. 
Dashed lines show the same probability distribution functions of the distances to the first 4 nearest waters when the water molecules were randomly selected from the whole ensemble. We select the same number of random waters in the system as there are large jumping water molecules within the same time window.
For the cumulative distribution functions of distances to the nearest jumping molecules see Fig.S4.
}
\label{fig:clusters}
\end{figure*} 

To address this question, we analyzed the spatial position of the jumping molecules with respect to each other by calculating the distances to the 1st, 2nd, 3rd, and 4th nearest molecule that jumps within the selected time interval. We obtain the probability distribution function (PDF) for these distances by repeating the analysis for the time interval of 200 fs shifting the time window along the whole duration of the simulation. The obtained PDFs for the position of the nearest jumping molecules with respect to another jumping molecule are shown in Fig.\ref{fig:clusters}, when looking at the angular jumps of magnitude larger than $60^{\circ}$ and $40^{\circ}$, respectively. 
To understand whether the jumps are distributed homogeneously in space or not, we compare these PDFs to the case where, instead of the concurrently jumping molecules, we look at the same number of randomly selected molecules in the network. The PDFs for the distances of the first 4 nearest random molecules to another random molecule are shown by dashed lines in Fig.\ref{fig:clusters}. 

For angular jumps with $\Delta\Theta>60^{\circ}$, the difference in the PDFs is most evident for the 1st and 2nd nearest neighbor (panels A and B in Fig.\ref{fig:clusters}). The nearest jumping molecule is typically situated in the first water shell of another molecule that performs large jump, while the 2nd nearest neighbor has a wide distribution around the second water shell distance. Instead, if the concurrently jumping molecules would be distributed homogeneously throughout the system, as is the case when choosing random molecules, the two nearest neighbors would have been positioned at larger distances with a rather different shape of the PDFs (dashed black lines in Fig.\ref{fig:clusters}A and B). This gives a strong indication that the simultaneous large jumps we observe are not happening independently throughout the system, but are instead interconnected and correlated. 

Large jumps are usually preceded and followed by changes in the local topology of the underlying H-bond network, resulting in subsequent spatially near by large water reorientation events. When we include the large jumps with a somewhat smaller amplitude, $\Delta\Theta>40^{\circ}$, we find that the differences in the PDFs are even more pronounced for the 3rd and 4th nearest neighbors which are situated much closer in space to the jumping molecules with respect to the homogeneous random case (see Fig.\ref{fig:clusters}C and D). This means that the large jumps generate a myriad of relatively large reorientations that occur in spatially organized clusters due to the local H-bond network rearrangements. 



\begin{figure*}[!htb]
\includegraphics[width=\textwidth]{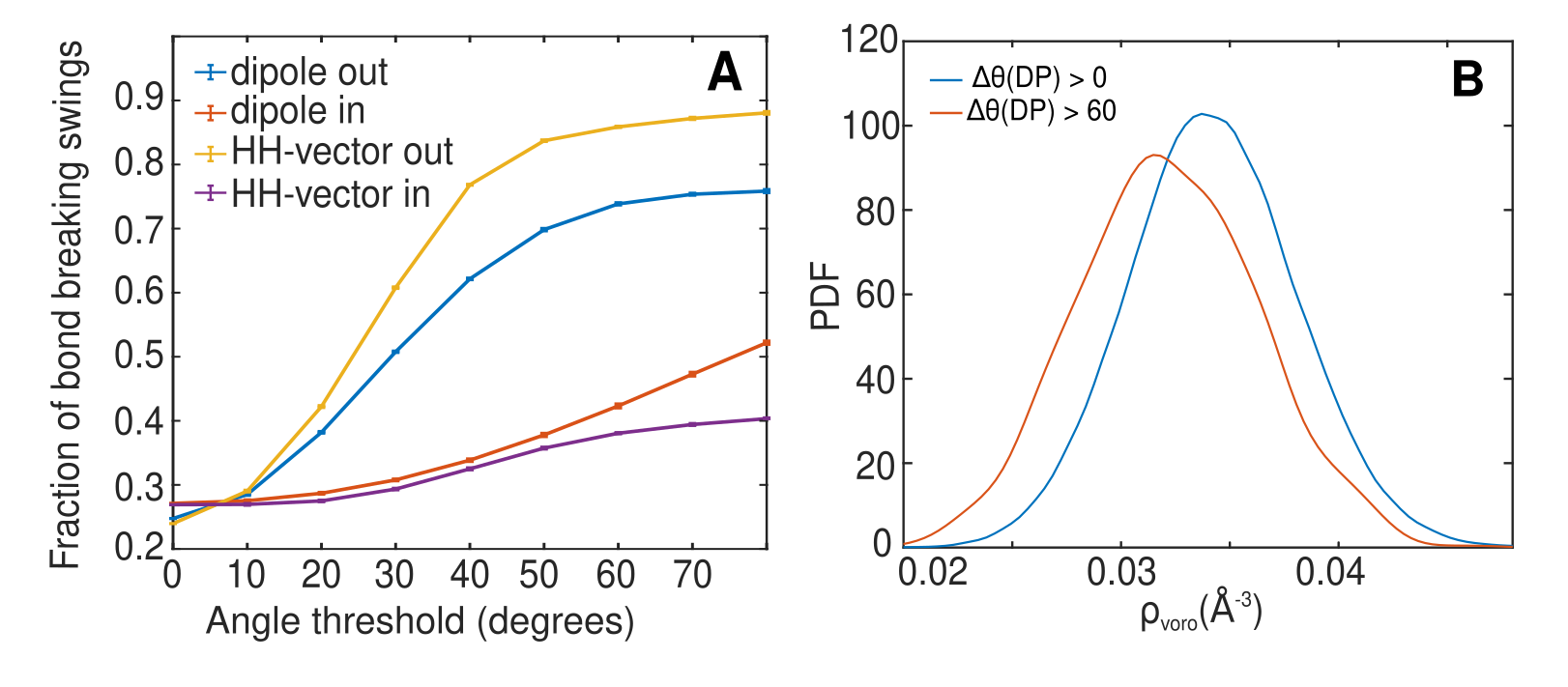}
\caption { Analysis of the local environment of large swings.
(A) The fraction of H-bonds that break during angular swing events increases as the threshold on the magnitude of angular swing is increased. We performed detailed analysis by distinguishing the swings that break either outgoing H-bonds (interactions in which the jumping water donates a hydrogen atom to a neighboring acceptor) or ingoing H-bonds (interactions in which the jumping water accepts a hydrogen from a neighboring donor).
We find that the H-bonds associated with the hydrogen atom of the swinging molecule (outgoing blue and yellow) is affected the most by the large angular swing of the water. As before, we distinguish the swings detected from the DP and from the HH vector time series (see Fig.S5B for the total fraction of bond breaking swings).
(B) Probability distributions of the Voronoi density, $\rho_{\text{voro}}$, for the water molecules undergoing angular swings constrained by a magnitude threshold. The probability distributions shifts towards lower local densities as we restrict ourselves to swings with the larger angular magnitude.
}
\label{fig:environment}
\end{figure*}

With all this dynamical activity in the hydrogen bond network involving highly cooperative processes, we can now return back to how this complex behavior is reflected in changes in the local water structure. Fluctuations of water can involve changes in local topology \cite{blumberg1984connectivity}, coordination structure \cite{geiger1984structure} and density \cite{geiger1982low}. Laage and Hynes for example, showed that rotating water molecules break hydrogen bonds with overcoordinated neighbours and subsequently form interactions with undercoordinated water molecules \cite{laage2006molecular,laage&hynes}. At the same time, some studies have also shown that modulations in the local density can also affect the water dynamics \cite{malenkov2006structure}. Since our automated protocol detects angular swings without initial imposition of hydrogen bonding interactions, it is interesting to examine the correlation between the collective reorientational dynamics and changes in the number of hydrogen bonds. We next investigate how many of the detected angular swings are actually bond breaking events, where we count all the events which have a change in the H-bonded neighbors before and after a jump. 

 
Fig.\ref{fig:environment}A illustrates that as one increases our threshold on the magnitude of the angular swing, the probability that this incurs a hydrogen-bond breaking event is enhanced. This feature is observed for both the DP and HH vectors. Interestingly, there also appears to be an asymmetry in the behavior of outgoing (donating hydrogen bonds) versus incoming (hydrogen bonds being accepted by the jumping water). Large swings associated with the HH vector break donating H-bonds roughly $90\%$ of the time, while accepting H-bonds are broken $40\%$  of the time. In the case of the DP vector, this asymmetry persists, although to a lesser degree ($70\%$ for out and  $40\%$ in). The difference in behavior of large swings between donating versus accepting hydrogen bonds is essentially rooted in the position of the proton along the respective hydrogen bond. In the case of donating hydrogen bonds, angular swings require a large change in the O-H bond. On the other hand, for accepting hydrogen bonds, it is possible to make reorientational motions without necessarily disturbing the orientation of the O-H bond from another water molecule that is further away.

Indeed, this is also consistent with the original large jump mechanism proposed by Laage and Hynes, which involves an exchange of the H-bonds acceptors \cite{laage2006molecular,laage&hynes}. Here, however, by using the automatized detection of the angular swings, without a priori selecting the events based on the H-bond breaking, we are able to uncover a multitude of the swinging events occurring in the system that are interrelated by the underlying H-bond network rearrangement. Importantly, the asymmetric nature of the H-bond breaking events will be collectively manifested as an anisotropic diffusion of the topological defects.

Another interesting aspect of Fig.\ref{fig:environment}A is that it shows that a sizable fraction of large angular swings involve the breaking of hydrogen bonds that are donated to the water molecule under study. For the dipole vector, approximately 50\% of the large jumps involve hydrogen bonds being broken on the donating side. This feature is consistent with the collective nature of the swings elucidated earlier. The breaking of hydrogen bonds being donated to jumping waters implies that there are at least \emph{several pairs} of water molecules that undergo simultaneous reorientational motions.

Besides the changes in local coordination topology dictated by direction of the hydrogen bonds, density fluctuations are central to understanding both the thermodynamic and dynamical properties of water\cite{english2011density,chandlerhydrophobicity}. We speculated that regions of the hydrogen bond network which are lower in density might be more susceptible to reorientational motions. For all the identified jump time intervals, we select the mid-points and studied the statistics of the Voronoi density \cite{offei2021high,maiti_characterization_2013} ($\rho_{\text{voro}}$) defined as the inverse of the Voronoi volume taken as a sum of the oxygen and two hydrogens of the jumping waters. Indeed, Fig.\ref{fig:environment}B shows that as we increase the threshold of the magnitude of the angular jumps, we observe that the peak in the distribution of the Voronoi density shifts towards smaller values. This shows that large angular swings are enhanced in low density environments.

\section*{Conclusions}


\begin{figure*}[!htb]
\includegraphics[width=\textwidth]{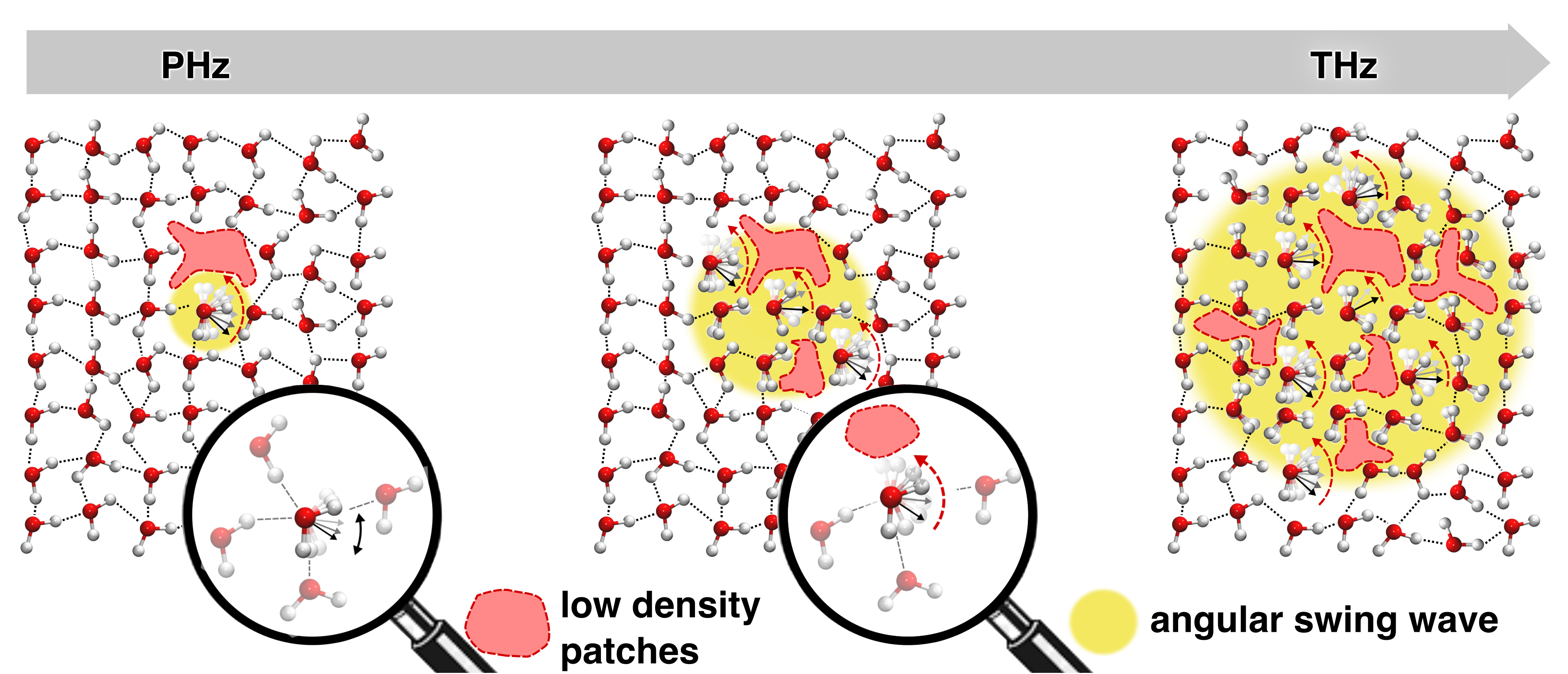}
\caption{The scheme depicts the essential ingredients associated with the collective angular jump mechanism in liquid water. Moving from left to right, large angular jumps near low-density regions of the water network, create a wave of jumps that propagates as depicted by the growing 
yellow
circle, to many different water molecules. Waters that are then close to low-density patches are also susceptible to large jumps. All in all, a highly collective process from the peta-to-teraHertz (PHz to THz) timescale occurs involving changes in topology and density triggered by the angular jumps.
}
\label{fig6_mech}
\end{figure*} 

In the present work, we elucidated the mechanism and origins of the collective reorientation in bulk water by employing molecular dynamics simulations and an automatized detection of abrupt angular fluctuations. The unsupervised protocol that detects large changes in the orientational motion of water molecules does not depend on any {\it a priori} criteria for hydrogen-bond interactions and allows identification of diverse non-local orientational swings. 

The picture that emerges from our communication is illustrated in the schematic shown in Fig.\ref{fig6_mech}. Essentially large angular jumps can occur at certain hot-spot regions, such as near low-density regions conferring greater rotational mobility to the water molecules. Due to the correlations in the hydrogen-bond network involving fluctuations in both topology and density, large jumps do not happen in isolation and subsequently, over the peta-to-TeraHertz timescale, we observe waves of both large and small jumps that happen concurrently. Although these dynamics have been observed in several previous spectroscopy based experiments \cite{fecko2005local,loparo2006multidimensional,laenen2002time}, the collective origin of these modes has remained poorly understood. 

One of the reasons why the interpretation of microscopic water dynamics is challenging even in the presence of large amount of simulation results is that the timescales of slow single-water reorientation processes are similar to those of the collective dynamics of many waters \cite{schulz2018collective}. 
While all the previous studies considered the concerted rotation of three water molecules while examining the reorientation dynamics, we have shown that the collective reorientation and H-bond network rearrangements involve several larger groups of near-by molecules that are spread throughout the system.  The spatial extent of the correlations and their coupling to thermal density fluctuations that occur on similar timescales  \cite{english2011density} would be interesting to study in the future.

Our results in bulk water at room temperature showing the highly cooperative character of the reorientational dynamics of the water molecules, open the doors to exploring how this effect changes upon supercooling and near biological systems where one might expect an enhancement of the phenomenon. Very recent results by Laage and coworkers suggest that the HB-breaking angular events are  related to water translational diffusion, pinpointing the connection between water's collective reorientation and its transport properties \cite{Gomez-JPCL-2022}.

Finally, our numerical results and molecular insights should motivate the creation of theoretical models to describe the cooperative dynamics in hydrogen bonded liquids \cite{perticaroli2017water} which may play an important role in tuning chemical reactions \cite{ruiz2020molecular}. The large angular swings we observe are akin to the tunneling dynamic pathways of water clusters at low temperature \cite{richardson2016concerted} which opens up interesting connections between the mechanisms of reorientational dynamics at low temperature and in the condensed phase.

\section*{Acknowledgments}
The authors acknowledge Francesco Paesani for sharing the trajectories of MB-pol water model.

\bibliographystyle{unsrt}  
\bibliography{references}

\end{document}